\documentstyle[11pt]{article}
\def\dalemb#1#2{{\vbox{\hrule height .#2pt
        \hbox{\vrule width.#2pt height#1pt \kern#1pt
                \vrule width.#2pt}
        \hrule height.#2pt}}}

\let\a=\alpha    \let\e=\epsilon

\def\nn{\nonumber} \def\bd{\begin{document}} \def\ed{\end{document}}
\def\ds{\documentstyle} \let\fr=\frac \let\bl=\bigl \let\br=\bigr
\let\Br=\Bigr \let\Bl=\Bigl 
\let\bm=\bibitem
\let\na=\nabla
\let\pa=\partial \let\ov=\overline
\def\ie{{\it i.e.\ }} 
\newcommand{\be}{\begin{equation}} 
\newcommand{\ee}{\end{equation}} 
\def\ba{\begin{array}}
\def\ea{\end{array}}
\def\ft#1#2{{\textstyle{{\scriptstyle #1}\over {\scriptstyle #2}}}}
\def\fft#1#2{{#1 \over #2}}
\def\del{\partial}
\def\sst#1{{\scriptscriptstyle #1}}
\def\oneone{\rlap 1\mkern4mu{\rm l}}
\def\e7{E_{7(+7)}}
\def\td{\tilde}
\def\wtd{\widetilde}
\def\im{{\rm i}}
\def\bog{Bogomol'nyi\ }
\newcommand{\ho}[1]{$\, ^{#1}$}
\newcommand{\hoch}[1]{$\, ^{#1}$}
\newcommand{\bea}{\begin{eqnarray}} 
\newcommand{\eea}{\end{eqnarray}} 
\newcommand{\ra}{\rightarrow}
\newcommand{\lra}{\longrightarrow}
\newcommand{\Lra}{\Leftrightarrow}
\newcommand{\ap}{\alpha^\prime}
\newcommand{\bp}{\tilde \beta^\prime}
\newcommand{\tr}{{\rm tr} }
\newcommand{\Tr}{{\rm Tr} } 
\newcommand{\NP}{Nucl. Phys. }
\newcommand{\tamphys}{\it Center for Theoretical Physics,
Texas A\&M University, College Station, Texas 77843}

\newcommand{\auth}{H. L\"u\hoch{\dagger}, S. Mukherji\hoch{\ddagger}, 
C.N. Pope\hoch{\dagger}}

\begin{document}
\begin{flushright}
\hfill{CTP TAMU-51/96}\\
\hfill{IC/96/211}\\
\hfill{hep-th/9610107}\\
\end{flushright}

\vspace{20pt}

\begin{center}
{\large {\bf Cosmological Solutions in String Theories}} 

\vspace{30pt}

\auth

\vspace{15pt}

{\tamphys}
\vspace{20pt}

K.-W. Xu

\vspace{15pt}

{\it International Center for Theoretical Physics, Trieste, Italy }

{\it  and }

{\it Institute of Modern Physics, Nanchang University, Nanchang, China }

\vspace{40pt}

\underline{ABSTRACT}
\end{center}

     We obtain a large class of cosmological solutions in the
toroidally-compactified low energy limits of string theories 
in $D$ dimensions.  We consider solutions where a $p$-dimensional
subset of the spatial coordinates, parameterising a flat space, a 
sphere, or an hyperboloid, describes the spatial sections of the
physically-observed universe.  The equations of motion reduce to 
Liouville or $SL(N+1,R)$ Toda equations, which are exactly solvable.
We study some of the cases in detail, and find that under suitable 
conditions they can describe four-dimensional expanding universes.
We discuss also how the solutions in $D$ dimensions behave upon
oxidation back to the $D=10$ string theory or $D=11$ M-theory.

{\vfill\leftline{}\vfill
\vskip	10pt
\footnoterule
{\footnotesize
	\hoch{\dagger}	Research supported in part by DOE 
grant DE-FG05-91-ER40633 \vskip	-12pt}  \vskip	10pt
{\footnotesize 
        \hoch{\ddagger} Research supported in part by NSF grant PHY-9411543}
}

\pagebreak
\setcounter{page}{1}

\section{Introduction}

     The study of cosmological consequences of string theory has been an
area of much active research in the past [1-5].  Recently, exploiting
certain duality symmetry, it has been suggested that inflation can occur in
string theory without relying on the potential energy density of the dilaton
field \cite{mggv,rbgv}. Since then, several papers have investigated aspects
of string-inspired cosmology in various dimensions [8-13]. One of the
intriguing features in these models is that a dynamical compactification of
some of the coordinate directions can occur, implying, for example, that a
cosmological solution of the ten-dimensional string can evolve so that it
describes a four-dimensional expanding universe, with the extra six
dimensions undergoing a contraction (see, for example, \cite{low}).  In view
of the recent advances in the understanding of the unity among string
theories, it can be argued that the most natural arenas within which
cosmological models should be studied are the type IIA and type IIB string
theories, or possibly their M-theory and F-theory progenitors \cite{low}.
Cosmological solutions using the R-R fields in type II string theories, as
well as those using the NS-NS fields, have been discussed recently in
\cite{low}.  In this paper, we shall examine a broad class of cosmological
models that arise as solutions of the low-energy limits of ten-dimensional
string theories or M-theory.   Our starting point will be to consider models
in the $D$-dimensional toroidal compactifications of the fundamental
theories, in which $p$-dimensional spatial sections (which may be flat,
spherical or hyperbolic) expand while $q=D-p-1$ dimensions contract, as the
universe evolves.  These models themselves may all be re-interpreted as
ten-dimensional or eleven-dimensional solutions, by reversing the reduction
procedure that led to the $D$-dimensional theory. By this means, large
classes of ten or eleven dimensional solutions of a cosmological type can be
obtained. The extra dimensions that are restored in this oxidation process
will themselves typically either expand or contract, adding further to the
numbers $p$ and $q$ of expanding and contracting dimensions.  Ultimately,
phenomenological considerations would imply that one is principally
interested in the case where the total number of expanding spatial
dimensions is 3. 

     The reason for dividing the discussion into two stages, in which first
we find cosmological solutions in the $D$-dimensional toroidal
compactifications of the ten-dimensional theories, and then we oxidise them
back to $D=10$ or $D=11$, is the following.  In order to construct
solutions, it is useful to choose a highly symmetrical ansatz for the form
of the metric and other fields in the theory, so as to obtain relatively
simple equations of motion for the remaining degrees of freedom.  Upon
oxidation to $D=10$ or $D=11$, the solutions typically acquire a more
complicated form, which would be less easily found, or classified, by a
direct ten-dimensional study of the equations of motion.  This is analogous
to the situation for $p$-brane solitons, where most of the lower-dimensional
solutions yield rather complicated configurations in $D=10$ or $D=11$ that
correspond to sets of intersecting $p$-branes, possibly also with ``boosts''
along certain directions in the toroidally-compactified dimensions 
[15-20]. 

     The organisation of the paper is as follows.  In section 2, we set up 
the ansatz for the metric tensor and the other fields of the theories, and 
then obtain cosmological models involving single-scalar, dyonic, and 
multi-scalar solutions.  Included in the multi-scalar cases are models where 
the equations of motion reduce to the $SL(N+1,R)$ Toda equations.  In 
section 3, we discuss the cosmological properties of some of the models.  In 
section 4, we discuss the Kaluza-Klein reduction and oxidation of the 
cosmological solutions.  The paper ends with conclusions in section 5.

\section{Cosmological solutions from string theories}

    Cosmological models are described by solutions of the low-energy 
effective theory in which the metric tensor, and the other fields, are time 
dependent.  We shall look for solutions in which the $D$-dimensional metric
takes the form 
\be 
ds^2 = -e^{2U}\, dt^2 + e^{2A} \, d\bar s^2 + e^{2B}\, dy^m\, dy^m\ ,
\label{metric}
\ee
where the functions $U$, $A$ and $B$ depend only on $t$, and $d\bar s^2$ 
represents the $p$-dimensional metric on the spatial section of a
$d$-dimensional spacetime, with $d=p+1$.   Typically, we shall consider
spatial metrics of the maximally-symmetric form 
\be
d\bar s^2 = \fft{dr^2}{1-k r^2} + r^2 d \Omega^2\ ,\label{max}
\ee
where $d\Omega^2$ is the metric on a unit $(d-2)$-sphere.  Without loss of 
generality, the constant $k$ may be taken to be equal to 0, 1 or $-1$, in 
which case metric $d\bar s^2$ describes flat, spherical, or hyperboloidal 
spatial sections respectively.  The general idea 
will be to look for solutions where this spacetime expands at large time,
while the $q=D-d$ dimensional space parameterised by the coordinates $y^m$
contracts and becomes unobservable at large time.  Note that the function
$U$ is redundant, in the sense that it can later be set to any desired 
form by an appropriate redefinition of the time coordinate.  It is 
convenient to include it, however, since the solution of the equations of 
motion can be simplified by choosing it appropriately. 

     In the vielbein basis $e^0= e^U\, dt$, $e^a=e^A\, \bar e^a$, $e^m= e^B\, 
dy^m$ (there will be no confusion between the exponential functions and the 
vielbeins), we find that the curvature 2-forms are given by
\bea
\Theta^0{}_a &=& e^{-2U}\, (\ddot A -\dot U \dot A + \dot A^2) e^0\wedge e^a
\ ,\nn\\
\Theta^0{}_m &=& e^{-2U} \, (\ddot B -\dot U \dot B +\dot B^2) e^0\wedge e^m
\ ,\nn\\
\Theta^a{}_b &=& \overline\Theta^a{}_b + e^{-2U} \dot A^2 e^a\wedge e^b\ ,
\label{2form}\\
\Theta^a{}_m &=& e^{-2U}\, \dot A \dot B e^a\wedge e^m\ ,\nn\\
\Theta^m{}_n &=& e^{-2U}\, \dot B^2 \, e^m\wedge e^n\ ,\nn
\eea
where a dot denote a derivative with respect to the time coordinate $t$, and
$\overline\Theta^a{}_b$ is the curvature 2-form for the metric $d\bar 
s^2=\bar e^a \bar e^a$ of the spatial sections, in the vielbein basis $\bar
e^a$.   It follows that the tangent-space components of the Ricci tensor for
the metric (\ref{metric}) are given by 
\bea 
R_{00} &=& -e^{-2U}\Big(p ( \ddot A + {\dot A}^2 -\dot U \dot A ) +q ( \ddot B 
+ {\dot B}^2 - \dot U \dot B ) \Big) \ ,\nonumber\\
R_{ab} &=& e^{-2U} ( \ddot A + p {\dot A}^2 - \dot U \dot A 
+ q \dot A \dot B ) {\delta}_{ab} + e^{-2A} \bar R_{ab} \ ,\label{ricci}\\
R_{mn} &=& e^{-2U} ( \ddot B + q {\dot B}^2 - \dot U \dot B
+ p \dot A \dot B ) {\delta}_{mn} \ .\nonumber
\eea
where $\bar R_{ab}$ denotes the tangent-space components of the Ricci tensor
for the spatial metric. In all the cases we shall consider, the metric
$d\bar s^2$ is Einstein, and we may write $\bar R_{ab} = k(p-1)
\delta_{ab}$, which is in particular the case for the metrics (\ref{max}). 

    In the rest of this section, we shall derive various classes of 
cosmological solutions in toroidally-compactified string theories.

\subsection{One-scalar cosmological solutions}

         The simplest cosmological solution in $D$ dimensions involves
the metric, a dilaton and an $n$-rank antisymmetric field strength $F_n$.
The Lagrangian is given by
\be
e^{-1}{\cal L} = R - \ft12 (\del \phi)^2 - \fft{1}{2n!}\, e^{a\phi}
\, F_n^2\ ,
\label{slag}
\ee
where the constant $a$ can be parameterised as \cite{lpss1}
\be
a^2 = \Delta - \fft{2(n-1)(D-n-1)}{D-2}\ .\label{avalue}
\ee
(Note that the Brans-Dicke Lagrangian generalised to include a field
strength can be recast in the form (\ref{slag}), by making a
dilaton-dependent conformal rescaling of the metric, and scaling the dilaton 
by an appropriate constant.  The Brans-Dicke parameter $\omega$ is related 
to the dilaton coupling constant $a$.)
In supergravity theories, the full bosonic Lagrangian can be consistently 
truncated to the single-scalar lagrangian (\ref{slag}) for $\Delta = 4/N$,
where $N$ is a set of integers $1,2,\ldots,N_{\rm max}$, where $N_{\rm max}$ 
depends on $D$ and $n$ \cite{lpsol}.   For 1-form field strengths, the
Lagrangian can also be consistently truncated to a set of $N$ 1-forms for
which $\Delta$ takes the values $24/(N(N+1)(N+2))$ \cite{lpsln}.   The
equations of motion from the Lagrangian (\ref{slag}) are 
\bea
\Box \phi &=& \fft{a}{2n!} \, e^{a \phi} \, F^2\ ,\nonumber\\
R_{\sst{MN}} &=& \ft12 \del_{\sst M} \phi \,\del_{\sst N} \phi
+S_{\sst{MN}}\ ,\label{seom}\\
\del_{\sst M_1}(e e^{a\phi} F^{\sst{M}_1\cdots \sst{M}_n}) &=&0\ ,\nn
\eea
where $S_{\sst{MN}}$ is a symmetric tensor given by
\be
S_{\sst{MN}} = \fft{1}{2(n-1)!} e^{a\phi} \Big(F_{\sst{MN}}^2 - 
\fft{n-1}{n(D-1)} F^2 g_{\sst{MN}}\Big) \ ,
\ee

      There are two types of ans\"atze for the field strength $F_n$ that are 
compatible with the symmetries of the metric (\ref{metric}), giving 
rise to elementary and solitonic cosmological solutions.  In the elementary 
solutions, the ansatz for the antisymmetric tensor is given in terms of its 
potential, and in a coordinate frame takes the form
\be
A_{m_1i_2\cdots m_q} = f \epsilon_{m_1m_2\cdots m_q}\ ,\label{eleans}
\ee
and hence
\be
F_{0m_1m_2\cdots m_q} = \dot f \epsilon_{m_1m_2 \cdots m_q}
\ ,\label{eleans1}
\ee
where $f$ is a function of $t$ only.  Here and throughout this paper 
$\epsilon_{\sst M\cdots \sst N}$ and $\epsilon^{\sst M \cdots \sst N}$
are taken to be tensor densities of weights $-1$ and $1$ 
respectively, with purely numerical components $\pm 1$ or $0$.  Note in 
particular that they are not related just by raising and lowering indices 
using the metric tensor.   For elementary solutions, we have $p=D-n$ and 
$q=n-1$.

      For the solitonic cosmological solutions, the ansatz for the 
tangent-space components for the antisymmetric tensor is
\be
F_{a_1 a_2\cdots a_p} = \lambda e^{-pA}\,\epsilon_{a_1a_2 \cdots a_p} \ ,
\label{solans}
\ee
where $\lambda$ is a constant.  Thus we have $p=n$ and $q=D-n-1$.  The form
of the exponential prefactor is determined by the requirement that $F_n$
satisfy the Bianchi identity $d F_n=0$. 

     Substituting the ans\"atze for the metric and the field strength
into the equations of motion (\ref{seom}), we find
\bea
\ddot \phi + (p\dot A + q\dot B -\dot U) \dot \phi &=& \ft12 \epsilon
a \lambda^2 \, e^{-\epsilon a\phi -2pA+2U}\ ,\nonumber\\
\ddot A + (p\dot A + q\dot B - \dot U) \dot A + k(p-1) e^{2U-2A}&=&
\fft{q}{2(D-2)} \lambda^2 \, e^{-\epsilon a\phi -2pA+2U}\ ,\nonumber\\
\ddot B + (p\dot A + q\dot B - \dot U) \dot B &=&
-\fft{p-1}{2(D-2)} \lambda^2 \, e^{-\epsilon a\phi -2pA+2U}\ ,\label{4eoms}\\
p(\ddot A + \dot A^2 - \dot U \dot A) +
q(\ddot B + \dot B^2 - \dot U \dot B) + \ft12 \dot \phi^2 &=& 
-\fft{p-1}{2(D-2)} \lambda^2 \, e^{-\epsilon a\phi -2pA+2U}\ ,\nonumber
\eea
where $\epsilon =1$ for the elementary case and $\epsilon=-1$ for the
solitonic case.  In the elementary case, the constant $\lambda$ arises as 
the integration constant for the function $f$ in (\ref{eleans}).

    It is convenient to make the gauge choice $U=p A + q B$, and to define
\be
X\equiv q B + (p-1)A\ ,\qquad Y\equiv B + \fft{(p-1)}{\epsilon a(D-2)}\, 
\phi\ ,\qquad \Phi\equiv -\epsilon a\phi +2q B\ .\label{redefs}
\ee
The equations of motion for $X$, $\Phi$ and $Y$ become 
\be
\ddot X+ k(p-1)^2 \, e^{2X}=0\ , \qquad \ddot\Phi +\ft12 \Delta \lambda^2\,
e^{\Phi}=0\ , \qquad\ddot Y =0\ ,\label{xypeq}
\ee 
together with the first integral
\be
\dot\Phi^2 +\Delta\lambda^2 e^{\Phi} +\fft{2q(D-2)a^2}{p-1}\dot Y^2 =
\fft{2p\Delta}{p-1} \Big(\dot X^2 + k(p-1)^2 e^{2X}\Big)\ .\label{cons}
\ee
Thus $X$ and $\Phi$ both satisfy Liouville equations.  The manifest
positivity of the left-hand side of (\ref{cons}) shows that the Hamiltonian
$\dot X^2 + k(p-1)^2e^{2X}$ for $X$ must be positive, and hence the
appropriate form of the solution is 
\bea
e^{-X} &=&\cases{\fft{p-1}{c}\, \cosh(ct +\delta)\ , & if $k=1$;\cr
\fft{p-1}{c}\, \sinh(ct +\delta)\ , & if $k=-1$;\cr}\nn\\
X&=& -ct-\delta\ ,\quad \hbox{if $k=0$}, \label{cases}
\eea
where $c$ and $\delta$ are constants. Note that in taking the square root of
$e^{2X}$, the positive root should be chosen in the expression for $e^{-X}$.
The Hamiltonian $\dot\Phi^2 + \Delta\lambda^2\, e^{\Phi}$ for $\Phi$ is also 
manifestly positive, and so the solution can be written as
\be
e^{-\ft12\Phi}= \fft{\lambda \sqrt\Delta}{2\beta}\, \cosh(\beta t + \gamma)
\ ,\label{psol}
\ee
where $\beta$ and $\gamma$ are constants.  The solution for $Y$ may be taken 
to be simply
\be
Y= -\mu t\ .\label{ysol}
\ee
The constraint (\ref{cons}) therefore implies that
\be
\beta^2= \fft{p\Delta c^2-q(D-2) a^2 \mu^2}{2(p-1)}\ .\label{beta}
\ee

     In terms of the original functions $A$, $B$ and $U$ appearing in the
metric (\ref{metric}), and the dilaton $\phi$, the solution takes the form 
\bea
e^{\ft{\Delta(D-2)}{2q} A} &=& \fft{\lambda \sqrt\Delta}{2\beta} \,
\cosh(\beta t +\gamma)\, 
e^{\ft{a^2(D-2)\mu t}{2(p-1)}}e^{\ft{\Delta(D-2)}{2q(p-1)} X} \ ,\nn\\
e^{-\ft{\Delta (D-2)}{2(p-1)} B} &=& \fft{\lambda \sqrt\Delta}{2\beta} \,
\cosh(\beta t +\gamma)\, e^{\ft{a^2(D-2)\mu t}{2(p-1)}}
\ ,\label{sssol}\\
e^{\ft{\Delta}{2\epsilon a}\phi} &=& \fft{\lambda \sqrt\Delta}{2\beta} \,
\cosh(\beta t +\gamma)\, e^{-\mu q t}\ ,\nn
\eea
together with $U= p A + q B$.  In a case where there is no dilaton, the 
solutions for $A$ and $B$ are again given by (\ref{sssol}), with $\mu=0$.
If instead $q=0$, we have from (\ref{avalue}) that $a^2=\Delta$; the 
solution for $\phi$ follows from (\ref{sssol}) by setting $q=0$, and $A$ is 
given by $A=X/(p-1)$, with $X$ given by (\ref{cases}).

\subsection{Dyonic cosmological solutions}

      In general dimensions, the cosmological solutions are either 
elementary or solitonic. In $D=2n$, the $n$-rank field strength can carry
both electric (\ref{eleans1}) and magnetic (\ref{solans}) charges.  In this
case, $p=n$ and $q=n-1$.  Making the same gauge choice $U=n A + (n-1) B$,
we find that the equations of motion are given by 
\bea
&&\ddot \phi = \ft12 a (\lambda_1^2 e^{-a\phi} - \lambda_2^2 e^{a\phi})
\, e^{2(n-1) B}\ ,\nonumber\\
&&\ddot A + k(n-1) e^{2(n-1)(A + B)} = \ft14 (\lambda_1^2 e^{-a\phi} + 
\lambda_2^2 e^{a\phi}) \, e^{2(n-1) B}\ ,\nonumber\\
&&\ddot B = -\ft14 (\lambda_1^2 e^{-a\phi} + \lambda_2^2 e^{a\phi} )
\, e^{2(n-1) B}\ ,\label{dyoneom}\\
&&p(\ddot A + \dot A^2 - \dot U \dot A) + q(\ddot B + \dot B^2 -
\dot U \dot B) + \ft12 \dot \phi^2 = -\ft14
(\lambda_1^2 e^{-a\phi} + \lambda_2^2 e^{a\phi} )
\, e^{2(n-1) B}\ .\nonumber
\eea
These equations can be simplified by defining the variables $X$,
$q_1$ and $q_2$:
\bea
X&=& (n-1) (A + B)\ ,\nonumber\\
B&=& \fft1{4(n-1)} \Big( q_2 + q_1 -2 \log((n-1)\lambda_1\lambda_2)\Big)\ ,\\
\phi &=& \fft{a}{2(n-1)} (q_2 - q_1) + \fft1{a}
\log\fft{\lambda_1}{\lambda_2}\ ,\nonumber
\eea
leading to
\be
\ddot X + k(n-1)^2 e^{2X} =0\ ,\qquad
\ddot q_1 = - e^{\a q_1 + (1-\a) q_2}\ ,\qquad
\ddot q_2 = - e^{(1-\a) q_1 + \a q_2}\ ,
\ee
together with the first-order constraint
\be
\ft12\a(\dot q_1^2 + \dot q_2^2) + (1-\a) \dot q_1 \dot q_2
+ e^{\a q_1 + (1-\a) q_2} + e^{\a q_2 + (1-\a) q_1} =
2n\Big( \dot X^2 +  k(n-1)^2 e^{2X} \Big)\ .
\ee
Here the constant $\a$ is given by
\be
\a = \ft12 + \fft{a^2}{2(n-1)} = \fft{\Delta}{2(n-1)}\ .
\ee
The solution for $X$ is straightforward, and is given by 
(\ref{cases}), with $p=n$.   For generic values of $\a$, a particular solution
for $q_1$ and $q_2$ can be obtained by setting $q_1=q_2$, which
reduces the equations to two identical Liouville equations.  This special
solution describes a self-dual cosmological model, with
$\lambda_1=\lambda_2$. It is unclear how to solve the equations in the
general dyonic case $\lambda_1\ne \lambda_2$,  for generic values of $\a$.  
However there are two values of $\a$ for which the equations are solvable.  
When $\a=1$, they reduce to two Liouville equations and the 
solutions can be straightforwardly obtained.  This value of $\a$ can arise 
for a 3-form field strength in $D=6$, with $\Delta=4$.   Another value of
$\a$ for which the equations are solvable is $\a=2$, in which case they
become the $SL(3,R)$ Toda equations.  This value of $\a$ can arise for a
2-form field strength in $D=4$, with $\Delta=4$.  (In fact, this field
strength can also support a dyonic black hole solution, whose equations of
motion can again be re-expressed as the same $SL(3,R)$ Toda equations
\cite{lpw1}.  This solution is unstable, in the sense that it is a bound
state of an electric and a magnetic black hole with negative binding energy
\cite{gk}.)

\subsection{Multi-charge cosmological solutions}

    The $D$-dimensional bosonic Lagrangian of M-theory to $D$ dimensions 
compactified on a torus can be consistently truncated to
\be
e^{-1} {\cal L} = R -\ft12 (\del \vec \phi)^2 -\fft1{2n!}
\sum_{\a=1}^{N} e^{\vec c_\a \cdot \vec \phi}\, F_\a^2\ ,\label{multilag}
\ee
when the dilaton vectors for the set of $N$ field strengths $F_\a$
of rank $n\ge 2$ satisfy the dot products
\be
M_{\a\beta} = \vec c_\a \cdot \vec c_\beta =
4 \delta_{\a\beta} - \fft{2(n-1)(D-n-1)}{D-2}\ .\label{mdot1}
\ee
The maximum value $N_{\rm max}$ for  $N$ depends on the rank of the field
strengths, and on the dimension $D$.  For example for 2-form field
strengths, $N_{\rm max}= 2$ for $6\le D \le 9$;  $N_{\rm max}=3$ in $D=5$;
and $N_{\rm max} = 4$ in $3\le D\le 4$ \cite{lpsol}.  We shall discuss the
case of 1-form field strengths in subsection 2.4.   In fact, we can perform
a further truncation to the single-scalar Lagrangian (\ref{slag}) with $a$,
$\phi$ and $F$ given by \cite{lpsol} 
\bea
a^2 &=& \Big(\sum_{\a,\beta} (M^{-1})_{\a\beta}\Big)^{-1}\ ,\qquad
\phi = a \sum_{\a,\beta} (M^{-1})_{\a\beta}\,\vec c_\a \cdot \vec\phi\ ,
\nonumber\\
F_\a^2 &=& a^2 \sum_\beta (M^{-1})_{\a\beta} F^2\ .\label{multitos}
\eea
For dilaton vectors whose dot products satisfy (\ref{mdot1}), the value of 
the constant $a$ is given by (\ref{avalue}) with $\Delta = 4/N$.

    In this subsection, we shall obtain multi-charge cosmological solutions 
for the Lagrangian (\ref{multilag}).   We use the same elementary
(\ref{eleans}) or solitonic (\ref{solans}) ans\"atze for the field strengths
$F_\a$.  The metric ansatz is given by (\ref{metric}), and again we make
the gauge choice $U=pA + qB$.  The equations of motion become
\bea
&&\ddot {\vec \phi} =\ft12 \epsilon \sum_\a \vec c_\a \, 
\lambda_\a^2\, e^{-\epsilon\vec c_\a\cdot\vec\phi + 2qB}\ ,\nn\\
&&\ddot A = -k(p-1) e^{2qB +2 (p-1) A}+\fft{q}{2(D-2)} \sum_\a
\lambda_\a^2\, e^{-\epsilon\vec c_\a\cdot\vec\phi + 2qB}\ ,\nn\\
&&\ddot B = -\fft{p-1}{2(D-2)} \sum_\a 
\lambda_\a^2\, e^{-\epsilon\vec c_\a\cdot\vec\phi + 2qB}
\ .\label{multieom1}\\
&&p(\ddot A + \dot A^2 -\dot U \dot A) +
q(\ddot B + \dot B^2 - \dot U \dot B) + \ft12 (\dot {\vec\phi})^2 =
-\fft{p-1}{2(D-2)} \sum_\a 
\lambda_\a^2\, e^{-\epsilon\vec c_\a\cdot\vec\phi + 2qB}\ .
\nonumber
\eea
As in the single-charge case, it is convenient to define new variables:
\be
X\equiv qB + (p-1) A\ ,\qquad Y = B + \fft{\epsilon (p-1)}{D-2} 
\sum_{\a,\beta} (M^{-1})_{\a\beta} \, \varphi_\beta\ ,\qquad
\Phi_\a = -\epsilon \varphi_\a + 2 q B\ ,
\ee
where $\varphi_\a = \vec c_\a\cdot \vec \phi$.   The equations of motion 
for $X$ and $\Phi_\a$ become a set of Liouville equations
\be
\ddot X + k(p-1)^2 e^{2X} = 0\ ,\qquad
\ddot \Phi_\a + 2\lambda_\a^2\, e^{\Phi_\a}=0\ ,
\ee
together with the first integral constraint
\be
\sum_\a \Big(\dot \Phi_\a^2 + 4\lambda_\a^2\,e^{\Phi_\a}\Big) + 
\fft{8q(D-2)a^2}{(p-1)\Delta} \dot Y^2 =
\fft{8p}{p-1} \Big(\dot X^2 + k(p-1)^2 e^{2X}\Big)\ ,\label{con2}
\ee
where $\Delta = 4/N$ and $a$ is given by (\ref{avalue}), and $Y$ again 
satisfies $\ddot Y=0$.

     The solution for the function $X$ depends on the value of the parameter 
$k$, and is again given by (\ref{cases}).  The solutions for $\Phi_\a$ take
the form 
\bea
e^{-\ft12 \Phi_\a} = \fft{\lambda_\a}{\beta_a} \cosh(\beta_\a t + 
\gamma_\a)\ ,
\eea
where $\beta_\a$ and $\gamma_\a$ are constants.  The solution for $Y$ may
again be taken to be simply $Y=-\mu t$.  The constraint (\ref{con2})
therefore implies that 
\bea
\sum_\a \beta_\a^2 = \fft{2p\Delta c^2- 2q(D-2)a^2\mu^2}{(p-1)\Delta}
\ .
\eea
In terms of the functions $U$, $A$, $B$ and the dilatonic fields 
$\varphi_\a$, the solutions can be expressed as
\bea
e^{-\ft{2(D-2)}{p-1} B} &=& e^{\ft{2(D-2)a^2 \mu t}{(p-1) \Delta}}
\prod_{\a=1}^{N} \Big(\fft{\lambda_\a}{\beta_\a} \cosh(\beta_\a t + 
\gamma_\a)\Big)\ ,\nonumber\\
e^{\ft{2(D-2)}q A} &=& e^{\ft{2(D-2)}{q(p-1)} X}
e^{\ft{2(D-2)a^2 \mu t}{(p-1) \Delta}}
\prod_{\a=1}^{N} \Big(\fft{\lambda_\a}{\beta_\a} \cosh(\beta_\a t + 
\gamma_\a)\Big)\ ,\label{mulitsol}
\eea
with $U=pA + qB$ and $\epsilon\varphi_\a = 2qB - \Phi_\a$.  When all the 
parameters $\lambda_\a$ are equal and all $\beta_\a$ are equal, multi-charge 
solutions reduce to the single-scalar solution given by (\ref{sssol})

\subsection{$SL(N+1, R)$ cosmological solutions}

      In this subsection, we discuss cosmological solutions with 1-form 
field strengths.  The bosonic Lagrangian of M-theory compactified to $D$
dimensions on a torus can be truncated to one involving $N\le N_{\rm max}$ 
1-form field strengths, of the form (\ref{multilag}).  Such
consistent truncations are possible when the dilaton vectors of the retained
field strengths satisfy (\ref{mdot1}).  In this case, we have $N_{\rm
max}=2$ for $7\le D \le 8$; $N_{\rm max} = 4$ for $5 \le D \le 6$; $N_{\rm
max}=7$ for $D=4$ and $N_{\rm max} =8$ for $D=3$. 

     For the 1-form field strengths, alternative consistent truncations are
possible in cases where the dot products of the dilaton vectors do not satisfy
(\ref{mdot1}), but instead satisfy \cite{lpsln} 
\be
M_{\a\beta} = 4 \delta_{\a\beta} - 2 \delta_{\a, \beta+1} -
2 \delta_{\a, \beta-1}\ .\label{cartan}
\ee
This is in fact twice the Cartan matrix for $SL(N+1,R)$, and consequently,
as we shall see, the equations of motion of the consistently-truncated
system (\ref{multilag}) can be cast into the form of the $SL(N+1, R)$ Toda
equations, which are exactly solvable.  The multi-scalar multi-charge system
can be further truncated to a single-scalar system (\ref{slag}) using
(\ref{multitos}); these solutions have \cite{lpsln} 
\be 
a^2 = \Delta = \fft{24}{N(N+1)(N+2)}\ .
\ee
It was shown in \cite{lpsln} that sets of dilaton vectors with dot products 
given by (\ref{cartan}) arise in all toroidally-compactified supergravities 
in all dimensions $D\le 9$.  In this case $N_{\rm max} = 10-D$.

      We shall first consider elementary cosmological solutions.  The metric
for the elementary case involving 1-form field strengths is given by 
\be
ds^2 = -e^{2U} dt^2 + e^{2A} d\bar s^2 \ ,\label{metric2}
\ee
where $d\bar s^2$ is again the metric on the spatial sections, typically 
taking the form (\ref{max}). The field strengths take the elementary form 
given by (\ref{eleans}).  It is convenient to make the gauge 
choice $U=(D-1) A$, which implies that the equations of motion become
\bea
&&\ddot \varphi_\a = \ft12 \sum_\beta M_{\a\beta}\,\lambda^2_\beta
\, e^{-\varphi_\beta}\ ,\qquad \ddot A + k(D-2) e^{2(D-2)A} =0\ ,
\nonumber\\
&&\sum_{\a,\beta} (M^{-1})_{\a\beta} \, \dot \varphi_\a \dot \varphi_\beta +
\sum_\a \lambda_\a^2\, e^{- \varphi_\a} = 2(D-1)(D-2) \Big(\dot A^2
+ k e^{2(D-2) A}\Big)\ .
\eea
Making the further redefinition $\Phi_\a = -2\sum_\beta
(M^{-1})_{\a\beta}\, \varphi_\beta$, we find that $\Phi_\a$ satisfies
\bea
&&\Phi_\a'' = -\lambda_\a^2 \exp(\ft12\sum_\beta M_{\a\beta} \Phi_\beta)\ ,
\label{presl}\\
&&\ft14 \sum_{\a,\beta} M_{\a\beta}\, \dot \Phi_\a\dot\Phi_\beta +
\sum_\a \lambda_\a^2 \exp(\ft12 \sum_\beta M_{\a\beta} \Phi_\beta)=
2(D-2)(D-1) \Big(\dot A^2 + k e^{2(D-2) A}\Big)\ .\nonumber
\eea
Finally, the redefinition $\Phi_\a = q_\a - 4\sum_\beta (M^{-1})_{\a\beta}
\log \lambda_\beta$ removes the charges from the equations, and hence from 
(\ref{cartan}) we see that the equations become
\bea
\ddot q_1 &=& -e^{2q_1 -q_2}\ ,\nonumber\\
\ddot q_2 &=& -e^{-q_1 + 2q_2 - q_3}\ ,\nonumber\\
\ddot q_3 &=& -e^{-q_2 + 2q_3 - q_4}\ ,\label{sl2ntoda}\\
&&\cdots\nonumber\\
\ddot q_{\sst N} &=& e^{-q_{\sst N-1} + 2 q_{\sst N}}\ .\nonumber
\eea
Thus the functions $q_\a$ satisfy the $SL(N+1,R)$ Toda equations, whilst 
$A$ satisfies the Liouville equation.  The solutions are subject to the 
first-order constraint in (\ref{presl}), which can be re-expressed as
\be
{\cal H} = 2(D-1)(D-2) \Big(\dot A^2 + k e^{2(D-2) A}\Big)\ ,\label{hcon1}
\ee
where ${\cal H}$ is the Hamiltonian for the Toda equations 
(\ref{sl2ntoda}), given by
\be
{\cal H} = \ft14 \sum_{\a,\beta} M_{\a\beta}\, \dot q_\a \dot q_\beta +
\sum_\a \exp(\ft12 \sum_{\beta} M_{\a\beta} q_\beta)\ .
\ee

     The solution of the Liouville equation for $A$, whose form depends on
the value of the constant $k$, is given by (\ref{cases}) with $p=D-1$,
where $X= (D-2) A$.  The general solution to the $SL(N+1, R)$ Toda
equations (\ref{sl2ntoda}) can be given compactly in the form \cite{a} 
\be
e^{-q_\a} = \sum_{k_1 < k_2\cdots < k_\a}^{N+1} f_{k_1}\cdots f_{k_\a}\,
\Delta^2(k_1, \ldots, k_\a)\, e^{(\mu_{k_1} +\cdots + \mu_{k_\a}) t}\ ,
\ee
where $\Delta^2(k_1, \cdots,k_\a) = \prod_{k_i<k_j} (\mu_{k_i} -
\mu_{k_j})^2$ is the square of the Vandermonde determinant, and $f_k$ and 
$\mu_k$ are arbitrary constants satisfying
\be
\prod_{k=1}^{N+1} f_k = \Delta^2(1,2,\cdots, N+1)\ ,\qquad
\sum_{k=1}^{N+1} \mu_{k} = 0\ .
\ee
The Hamiltonian, which is conserved, takes the value ${\cal H}
=\ft12 \sum_{\a=1}^{\sst N+1} \mu_\a^2$.  It follows from (\ref{hcon1}) that
the Hamiltonian constraint implies
\be
\sum_{k=1}^{N+1} \mu_k^2 = 4(D-1)(D-2) c^2\ .
\ee

     It is straightforward to generalise the above discussion to solitonic 
cosmological solutions, where we have $p=1$ and $q=D-2$.  The equations of 
motion again can be cast into the form of $SL(N+1, R)$ Toda equations, and 
hence we can obtain exact solutions.

\section{Cosmological characteristics of the solutions}

    The solutions that we have obtained in the previous section have metrics 
of the form (\ref{metric}) in which the scale factors $e^{A}$ and $e^{B}$ 
evolve in time.  In order to obtain realistic cosmological models, a first 
requirement is that the scale factor $e^{A}$ for the spatial sections of 
the universe should evolve from a small value at early times to a large
value at late times.  Furthermore, one would hope that the scale factor 
$e^{B}$ for the additional $q=D-p-1$ dimensions parameterised by the $y^m$
coordinates would become small at large times, so that the additional 
dimensions become unobservable.

     In order to discuss the evolution of the solutions, it is useful to 
introduce a comoving time coordinate.  In cases where there is no dilaton, 
such as solutions of $D=11$ supergravity or M-theory, the choice of metric 
is unambiguous, and the comoving time $\tau$ is given by $\tau = \int 
e^{U} dt$.  In cases where there is a dilaton, such as solutions of the type 
IIA string in $D=10$, there are two natural metrics that one might consider, 
namely the Einstein-frame metric and the string-frame metric.  The former 
is the one that we have been considering thus far in the paper, and the
latter is related to it by the conformal rescaling $ds^2_{\rm string} =
e^{\phi} ds^2$.  The comoving time coordinate depends on the choice of
metric. 

     We shall begin by considering the simpler case of solutions in $D=11$ 
supergravity, and then afterwards we shall consider solutions in $D=10$.

\subsection{Cosmology in $D=11$}

      In $D=11$ supergravity, the bosonic fields consist only of the metric 
and a 4-form field strength.  We can use it to construct both elementary 
cosmological solutions with $p=7$ and solitonic solutions with $p=4$.  Their 
metrics are given by the first two equations in (\ref{sssol}), with
$\Delta=4$ and $\mu=0$, owing to the absence of the dilaton.  We may also
take $\gamma=0$, by appropriate choice of the origin for $t$.  Let us
consider the solitonic case, which will correspond to a 5-dimensional
cosmological model.  We see from (\ref{beta}) that $\beta^2=\ft83 c^2$ in
this case, and from (\ref{sssol}) that 
\bea
e^{3A} &=&\fft{\lambda}{\beta} \,e^X\, \cosh\beta t\ ,\nn\\
 e^{-6B}&=& \fft{\lambda}{\beta} \, \cosh\beta t\ ,\label{d11sol}
\eea
and $U=4A+6B$.  Without loss of generality, we may take $\beta$ to be
positive.  The scale factor $R\equiv e^A$ diverges both at $t=\infty$ and at
$t=-\infty$, since $\beta > |c|$, while the scale factor $e^B$ associated
with the extra $q$ dimensions tends to zero in both limits.  The comoving
time $\tau$, defined such that $ds^2 = -d\tau^2 + e^{2A} d\bar s^2 + e^{2B}
dy^m \, dy^m$, is given by $\tau=\int e^U\, dt$.   In general, owing to the
complexity of the function $e^{U}$, the relation between $\tau$ and $t$ can
only be evaluated by numerical methods. 

     If $k=1$, the comoving coordinate $\tau$ is finite for all values of
$t$.  When $t$ runs from $-\infty$ to $+\infty$, $\tau$ runs between two
finite values, $\tau_{-}$ to $\tau_{+}$.  In this process, the scale size
$R=e^{A}$ shrinks from infinite size at $\tau_{-}$ to a minimum at some
value $\tau_0$ and then expands again to infinity at $\tau_{+}$.  From
(\ref{2form}), and the form of the solution (\ref{d11sol}), we see that the
curvature is of order $e^{-2U}$ at large $|t|$, and thus diverges as $\exp(
(\ft83 -\ft23 \sqrt{\ft83})|ct|)$.  This solution is obviously undesirable
from both the phenomenological and the theoretical points of view.  The
situation is different when $k=0$, since then we have $e^{U}\sim
\exp(\ft13(\beta|t| -4 c t))$ at large $|t|$.  If $c$ is negative,
$c=-\sqrt{\ft38}\beta$, then $\tau$ diverges as $t$ tends to infinity, and
in fact $\tau \sim e^{U}$.  Since $e^{-2U}$ goes to zero at large $\tau$, it
follows that the curvature goes to zero at large $\tau$. Finally, if $k=-1$
the coordinate $t$ runs from $-\delta/c$ to $\pm\infty$, and $\tau$
correspondingly runs from infinity to zero.  The scale size $R$ becomes
large as $\tau$ tends to infinity, and is zero when $\tau$ is zero. The
curvature is singular at $\tau=0$, and tends to zero as $\tau$ tends to
infinity.  Thus both the $k=0$ and $k=-1$ models have the feature that the
universe expands as the comoving time increases from some finite time
$\tau_0$ to infinity, {\it i.e.}\ $R'>0$ for $\tau>\tau_0$, where a prime
denotes a derivative with respect to $\tau$. One may define $\tau_0$, where
the scale factor $R$ is a minimum, as the starting point for the expansion
of the universe.  In the $k=0$ case, we have $R'' >0$ at $\tau =\tau_0$ and
$R''<0$ as $\tau\rightarrow\infty$. When $k=-1$, we have $R''>0$ at
$\tau=\tau_0$, and $R''=0$ as $\tau$ tends to infinity.  In these $k=0$ and
$k=-1$ solutions, the universe is not starting from zero size at
$\tau=\tau_0$, but rather, this value of the comoving time represents the
point at which it has a minimum size, which is of the order of the Planck
scale.  In principle, one can extrapolate back to $\tau=0$, at which point
the scale size $R$ is infinite.  Although this region $0\le\tau\le\tau_0$
does not itself describe a satisfactory cosmological evolution, since the
comoving time reaches an endpoint at $\tau=0$ and the curvature diverges 
there, there is a sense in which one can think of the ``physical'' universe
with $\tau\ge\tau_0$ as emerging through a wormhole at $\tau=\tau_0$. 

     In a similar manner, one can analyse the elementary solutions in 
$D=11$, which describe an 8-dimensional cosmological model.

\subsection{Cosmology in $D=10$}

     Since ten-dimensional string theories have 3-form field strengths, we 
can obtain solitonic solutions with $p=3$ and $q=6$, which describe 
4-dimensional cosmological models.  The dilaton coupling is such that
$a^2=1$, and hence $\Delta=4$.  The string coupling constant is given by 
$g=e^{-\phi}$.  The NS-NS 3-form field strengths of any of the $D=10$ string
theories have dilaton coupling $a=+1$, while the R-R 3-form of the type IIB 
theory has $a=-1$.    From (\ref{sssol}), we find that the
solutions have the form
\bea
e^{\ft83 A} &=& \fft{\lambda}{\beta}\, \cosh(\beta t+\gamma)\, 
e^{2\mu t + \ft43 X}\ ,\nn\\
e^{-8B} &=& \fft{\lambda}{\beta}\, \cosh(\beta t+\gamma)\, e^{2\mu t}\ ,
\label{d10sol}\\
e^{-\ft{2}{a} \phi} &=& \fft{\lambda}{\beta}\, \cosh(\beta t+\gamma) \,
e^{-6\mu t}\ ,\nonumber
\eea
together with $U= 3A + 6B$, and from (\ref{beta}) we have $\beta^2 = 3c^2
-12 \mu^2$. 

      Let us first consider the case when $\mu=0$.  This value has the
distinguishing feature that the dimensional reduction of the solution by
compactifying the $y^m$ coordinates gives rise to solutions which also
involve only one scalar field, as we shall discuss in section 4. In the
Einstein frame, the analysis of the cosmological properties of these
solutions is analogous to that for $D=11$.  When $k=1$, the comoving
coordinate $\tau$ runs from $\tau_{-}$ to $\tau_{+}$ as $t$ runs from
$-\infty$ to $+\infty$.   The 4-dimensional scale size shrinks from infinity
to a minimum and then expands to infinity again. 
For the case $k=0$, and $c<0$, the comoving coordinate runs from
zero to infinity as $t$ runs from $-\infty$ to $\infty$.  The 
scale parameter $R=e^{A}$ diverges in both the $\tau\rightarrow 0$ and 
$\tau\rightarrow \infty$ regimes.  Thus we can define $\tau_0$, at which
the scale parameter $R$ is a minimum, as the starting point of the expansion
of the universe, with $\tau$ running from $\tau_0$ to infinity.  It is easy
to verify that speed of the expansion $R'$ is always greater than zero when
$\tau >\tau_0$, but with $R''>0$ when $\tau \rightarrow \tau_0$ and $R''<0$
when $\tau \rightarrow \infty$.   Although the metrics behave identically
for both the NS-NS and R-R solutions, the dilaton field, and hence the
string coupling $g$, behave in opposite ways.  For the NS-NS solution, the
string coupling diverges when $\tau\rightarrow \infty$, whilst for the R-R
solution the string coupling vanishes in that limit.   In all cases, the 
curvature tends to zero when the scale factor $R$ is large, if this
coincides with $\tau$ going to infinity, namely in the $k=0$ and $k=-1$
models. On the other hand, if large $R$ corresponds to a finite value of
$\tau$, as in the $k=1$ models, the curvature diverges there. 

      If $k=-1$, the coordinate $t$ runs from $\pm \infty$ to $t=-\delta/c$,
and correspondingly, the comoving coordinate $\tau$ runs from zero to
infinity.  The scale parameter $R$ runs from infinity to a minimum at
$\tau_0$, and then to infinity again.  Thus we can define the expansion of
the universe from $\tau=\tau_0$ to $\tau=\infty$.  At the beginning of the
universe, we have $R'' >0$ whilst at the end of universe we have $R''=0$ in
this case.   At the beginning $\tau=\tau_0$, the string coupling constant
$g$ is a non-vanishing constant for both NS-NS solutions and R-R solutions
and it converges to another non-vanishing constant at large $\tau$. 

       In the above discussion, we studied the cosmological characteristics 
of the metrics in the Einstein frame.  In this frame, the form of the 
metrics is insensitive to whether the solution is constructed using an
NS-NS 3-form or an R-R 3-form.  However, since the constant $a$ in the
dilaton prefactor is $+1$ for the NS-NS solutions and $-1$ for the R-R
solutions, the string metrics for the NS-NS and R-R cases are quite
different, and we shall discuss them separately. In the string-frame metric,
the scale factor $e^{B}$ for the $y^m$ space in the R-R solutions diverges
when $|t|$ goes to infinity.   Thus the solutions only make sense when
$k=-1$, since in this case the relevant part of the evolution does not
involve the large $|t|$ regime.  Specifically, as $t$ runs from $\pm\infty$
to $-\delta/c$, the comoving coordinate $\tau$ runs from zero to infinity,
but the evolution is taken from $\tau=\tau_0$ where $R=e^A$ is a minimum to
$\tau=\infty$ where $R$ diverges.  This behaviour of the scale factor $R$ is
similar to that in the Einstein frame, but the scale factor $e^{B}$ shrinks,
although remaining finite and non-zero for the entire evolution.  For the
NS-NS solutions, the string coupling diverges at large $|t|$.  Thus the
solutions are again restricted to the case $k=-1$, where $t$ runs from $\pm
\infty$ to $t=-\delta/c$.  The comoving coordinate $\tau$ runs from zero to
infinity.  At large $\tau$, the $D=4$ universe expands with constant speed;
but at $\tau=0$, unlike in the Einstein frame, we have $R=e^{A}=0$, and
$R''<0$ for small $\tau$. 

     So far, we have discussed the cosmological characteristics of the
solution (\ref{d10sol}) when $\mu$ is set to zero.   We saw that the string
coupling diverges for NS-NS solutions at large $|t|$, while it vanishes for
R-R solutions.  Now let us examine the solutions when $\mu$ is
non-vanishing.  In this case, as we shall see in the next section,
dimensional reductions of the solutions in which the $y^m$ coordinates are
compactified give rise to solutions with additional scalar fields in the
lower dimension. We shall examine the metrics for solutions with $\beta=6\mu
>0$. In this case the dilaton, and hence the string coupling, becomes a
constant when $t$ goes to infinity, for both the NS-NS and the R-R
solutions.  Then, by making a small perturbation away from $\beta=6\mu$,
which will not qualitatively affect the characteristics of the metric, we
can have the string coupling vanish when $t$ goes to infinity.  It follows
from the equation below (\ref{d10sol}) that we have $|c|=4\mu$ when $\beta
=6\mu$. In the Einstein frame, the behaviour of the solutions for $k=1$,
$k=0$ and $k=-1$ are analogous to the corresponding ones with $\mu=0$ that
we discussed previously.  This implies, in particular, that by adjusting the
parameter $\mu$ properly, we can have an inflationary model even for NS-NS
solutions when $k=0$, where the string coupling vanishes as the comoving
time approaches infinity.  At large $t$ the behaviour of the metric in the
string frame is the same as that in the Einstein frame, since the dilaton
tends to a constant at large $t$.   If $k=0$, large values of $t$ imply
large values of $\tau$, whilst if $k=-1$, they imply that $\tau$ tends to
zero.  Thus in the string frame, by adjusting the parameter $\mu$ properly,
we can also have inflationary models where the string coupling respectively
vanishes or goes to a constant at large $\tau$, while the expansion rate
$R'$ of the 4-dimensional universe either tends to zero or becomes a
constant. 

     We have discussed the cosmological features of the solitonic
solutions for both the NS-NS and R-R 3-forms.   These solutions can provide
inflationary models of the universe.  The 9-dimensional space divides into
two parts:  a 6-dimensional subspace shrinks to zero or a finite size as the
comoving time tends to infinity, while a 3-dimensional subspace expands. The
solutions provide a dynamical compactification of the 10-dimensional
spacetime to $D=4$. 

     In $D=10$, there exist further field strengths of other ranks, and
the associated solutions will describe cosmologies in different dimensions. 

\section{Dimensional reduction and oxidation of cosmological solutions}

    In the previous sections, we constructed rather general classes of 
cosmological solutions in $D$-dimensional supergravity theories.  
Ultimately, one views these theories as originating from some fundamental 
theory such a string in $D=10$, or M-theory in $D=11$.  Since the 
lower-dimensional theories that we have considered are obtained by 
consistent dimensional reduction from $D=10$ or $D=11$, it follows that all 
their cosmological solutions can be oxidised back to solutions in the 
fundamental higher dimension.  In part, the utility of constructing 
solutions first in the lower dimension is that it can often be simpler than 
solving the equations directly in the fundamental dimension.  In particular, 
this is true if the lower-dimensional solution involves more than one field 
strength, since its oxidation to the higher dimension will then give a 
solution that lies outside the class that we have considered thus far.  Thus 
it is useful to study the general procedure of oxidation and reduction of 
the various cosmological solutions. 

\subsection{Kaluza-Klein dimensional reduction}

    The general procedure of toroidal dimensional reduction to can be broken 
down into a sequence of one-step reductions on circles.  The necessary 
reduction formulae can thus be encapsulated in the reduction of the 
following Lagrangian in $(D+1)$ dimensions,
\be
{\cal L}_{\sst D+1} = \hat e \hat R - \ft12 \hat e (\del\hat \phi)^2 
-\fft{e}{2 n!}\, e^{\hat a\hat\phi}\, \hat F_n^2\ ,\label{ddlag}
\ee
giving
\bea
{\cal L}_{\sst D} &=& e R -\ft12 e (\del\phi)^2 - -\ft12 e (\del\varphi)^2 
-\ft14 e \, 
e^{-2(D-1)\a\varphi} \, {\cal F}^2 \nn\\
&&-\fft{e}{2 n!}\, e^{-2(n-1)\a\varphi -\hat a\phi} \, {F'_n}^2 -
\fft{e}{2(n-1)!}\, e^{2(D-n)\a\varphi -\hat a \phi}\, F_{n-1}^2\ ,
\label{dlag}
\eea
in $D$ dimensions.  The $(D+1)$-dimensional hatted fields are expressed in 
terms of the $D$-dimensional unhatted fields by the standard Kaluza-Klein 
relations:
\bea
&&d\hat s_{\sst D+1}^2 = e^{2\a\varphi}\, ds_{\sst D}^2 +
e^{-2(D-2)\a\varphi}\, (dz + {\cal A})^2 \ ,\nn\\ 
&&\hat A_{n-1} = A_{n-1} + A_{n-2}\wedge dz\ ,\qquad \hat \phi=\phi\ ,
\label{kkans}
\eea
where all the unhatted fields are independent of the compactification 
coordinate $z$, and ${\cal F}=d{\cal A}$.  The constant $\a$ is given by
$\a=(2(D-1)(D-2))^{-1/2}$. The lower-dimensional field strengths, obtained
from the exterior derivative of the expression for the potential $\hat
A_{n-1}$ given above, are therefore expressed as $\hat F_n = F'_n +
F_{n-1}\wedge (dz +{\cal A})$, where $F'_n$ is the Chern-simons corrected
form $F'_n= d A_{n-1} -d A_{n-2} \wedge {\cal A}$, and $F_{n-1}= d A_{n-2}$.

      Let us begin by applying the above formalism to the example of a 
$D=10$ cosmological solution, of the form (\ref{sssol}) with $p=3$ and 
$q=6$, with $\Delta=4$ and $a^2=1$.  We shall take the compactification 
coordinate $z$ to be one of the $y^m$ coordinates, so that in $D=9$ we
have $p=3$ and $q=5$.  The relevant part of the $D=9$ Lagrangian will be, 
using (\ref{dlag}),
\bea
{\cal L}_9 &=& e R -\ft12 e (\del\phi)^2 -\ft12 e (\del\varphi)^2 
-\ft1{12}e\, e^{a\phi-4\a\varphi}\, F_3^2 \nn\\
&=&e R -\ft12 e (\del\phi_1)^2 -\ft12 e (\del\phi_2)^2 - \ft1{12} e\,
e^{b \phi_1} \, F_3^2 \ ,\label{d9lag}
\eea
where in the second line we have introduced a rotated pair of dilatonic 
scalars, defined by $b \phi_1= a \phi -4\a \varphi$ and $b \phi_2 = 4\a \phi 
+ a \varphi$. The constants $\a$ and $b$ are given by $\a=1/(4\sqrt7)$ and 
$b^2=8/7$.  Comparing the metric $ds^2_{10}$ given in (\ref{metric}) for the
ten-dimensional solution with the dimensionally-reduced metric $ds^2_9$ 
defined in (\ref{kkans}), we see that the Kaluza-Klein scalar $\varphi$ is 
given by $\a\varphi=-B/7$.   Thus it follows from the ten-dimensional 
solution (\ref{sssol}) that its dimensional reduction to $D=9$ gives a 
metric of the form (\ref{metric}), with functions $\tilde U$, $\tilde B$ and 
$\tilde B$ (the tildes denoting $D=9$ quantities) given by
\bea
e^{\fft{14}{5} \tilde A} &=& \fft{\lambda}{\beta}\, \cosh(\beta t +\gamma)
\, e^{2\mu t +\fft75 X}\ ,\nn\\
e^{-7\tilde B} &=& \fft{\lambda}{\beta}\, \cosh(\beta t+\gamma) \, 
e^{2\mu t} \ ,\label{d9sol}
\eea
together with $\tilde U= 2 \tilde A + 5 \tilde B$. The $D=9$ dilatonic
fields $\phi_1$ and $\phi_2$ are given by 
\bea
e^{-\ft2b \phi_1} &=& \fft{\lambda}{\beta}\, \cosh(\beta t+\gamma)\,
e^{-5\mu t}\ ,\nn\\
e^{-\ft{4\sqrt2}{b}\phi_2} &=& e^{-8\mu t}\ .
\eea
Thus we see that in general, the dimensional reduction of a single-scalar 
solution has given rise to a solution with two linearly-independent scalars. 
If, however, we consider the $D=10$ solution with $\mu=0$, then it reduces 
to a single-scalar solution in $D=9$.  In fact this $D=9$ solution is 
precisely of the same form (\ref{sssol}), with $p=3$, $q=5$, $\Delta=4$
and $\mu=0$.  It is interesting to note that even when $\mu$ is taken to be
non-zero, the $D=9$ metric is still of the form given in (\ref{sssol}).  
However, the Hamiltonian relation (\ref{beta}) between the integration
constants $\beta$, $c$ and $\mu$ is given by the ten-dimensional formula
$\beta^2 = 3c^2 -12\mu^2$ rather than the nine-dimensional formula
$\beta^2=3c^2 -10\mu^2$ that would be needed if the $\mu\ne0$ solutions were
to have the single-scalar form (\ref{sssol}) in $D=9$.  Thus the  reason why
the additional scalar $\phi_2$ is excited in the $\mu\ne0$
dimensionally-reduced solutions is that its energy contribution is needed in
order to make up the deficit in the Hamiltonian constraint. 

     It is interesting to note that the dimensional reduction of the
10-dimensional cosmological solutions gives rise to new solutions that are
beyond the scope of section 2, in that an additional scalar, namely
$\phi_2$, which does not couple to the field strength, becomes linearly
proportional to the time coordinate $t$.  The metric and the dilaton
$\phi_1$ of the solution, however, have exactly the same form as those for
the solution with vanishing $\phi_2$. In fact the constant of
proportionality $\nu$ in the time dependence of $\phi_2=\nu t$ can be
arbitrary, and has the effect of changing the relation between the constants
of integration to $\beta^2 =3c^2 - 10 \mu^2 -\nu^2$, where $\nu$ is the
contribution from $\phi_2$.   These $D=9$ solutions can also be oxidised to
$D=10$, but for generic values of $\nu$ the internal coordinate $z$ can no
longer be isotropically grouped with the coordinates $y^m$.  Thus these
$D=9$ solutions gives rise to new ten-dimensional cosmological solutions,
but in this case the $y^m$ and $z$ spaces dynamically compactify at
different rates. 

     The above illustration of Kaluza-Klein dimensional reduction of the
$D=10$ cosmological solutions to $D=9$ can be easily generalised to
arbitrary dimensions.   The reverse of the procedure provides a mechanism
for oxidising all the lower-dimensional solutions back to $D=10$ or $D=11$. 
Thus all the lower-dimensional solutions we obtained in section 2 can be
viewed as ten or eleven dimensional solutions, and obtaining such
lower-dimensional solutions provides a convenient algorithm for constructing 
and classifying sets of ten or eleven dimensional theories.  Of course it is
not guaranteed that the coordinate directions that are selected for this
non-dynamical Kaluza-Klein compactification of the theory will actually
shrink, rather than grow, as the cosmological solution evolves.  Which of
these occurs is a matter of calculation in the specific model in question. 
If it should turn out that some of the compactification directions actually
expand with time, it becomes natural, from the ten-dimensional point of
view, to include them in the spatial directions of the expanding universe.
For example, we can construct a $p=2$ solitonic solution in four dimensions
using a 2-form field strength.  From the four-dimensional point of view, we
have a three-dimensional expanding universe, with one shrinking circle.
However, if the 2-form field strength comes from the dimensional reduction
of the 3-form in $D=10$, oxidation of the four-dimensional solution to
$D=10$ reveals that one of the six ``compactifying'' coordinates in fact
expands, and hence gives rise to a four-dimensional expanding universe,
which is no different from the $p=3$, $q=6$ solutions with $k=0$ in $D=10$
that we discussed earlier. 

       However, this does not imply that all the lower-dimensional solutions 
are nothing but reductions of already-known higher-dimensional solutions. 
As we saw earlier, the single-scalar $D=9$ solution with non-vanishing
$\mu$ oxidises to a solution in $D=10$ that is not encompassed by the ansatz
in section 2.  In fact as we saw in section 2, a large number of solutions
arise in lower dimensions that involve more than one field strength.  The
equations of motion have the form of a set of Liouville equations or
$SL(N+1,R)$ Toda equations. The oxidation of these solutions provides a rich
variety of cosmological solutions in $D=10$.  It would be very interesting
to analyse their cosmological significance. 

\section{Conclusions}

    In this paper we have made a rather extensive study of certain classes 
of cosmological solutions in $D=10$ string theory or M-theory.  In 
particular, we began by constructing cosmological models in the 
$D$-dimensional toroidal compactifications of the string or M-theory, in
which the metric takes the form (\ref{metric}).  The relevant cosmological
solutions correspond to cases where $p$-dimensional spatial sections that
can be flat, spherical, or hyperboloidal expand in time, while an internal
$q$-dimensional space undergoes a contraction.   This spacetime, with 
$D=p+q+1$ dimensions, can then be embedded in the original $D=10$ or $D=11$ 
theory by reversing the steps of the toroidal compactification to $D$ 
dimensions.  This gives further dimensions that may be expanding or 
contracting, depending on the details of the solution.  The models that 
would be of principal interest for cosmology are those where the total 
number of expanding spatial dimensions is 3.  By dividing the process of 
constructing $D=10$ or $D=11$ solutions into these two stages, one can 
obtain rather broad classes of solutions with relative ease. 

    We examined some general features of the evolution of the metric scaling 
functions in some of the simpler solutions that we obtained.  In certain 
cases, we found that the behaviour of the scale parameters was of the 
phenomenologically-desirable form, in which the ``physical'' spatial 
dimensions grow from a very small initial size to a large size at later 
times, while the additional ``internal'' dimensions shrink, or dynamically 
compactify.  In particular, this kind of behaviour can arise in the $k=0$ 
and $k=-1$ models, where we find that the scale parameter $R$ of the 
physical spatial sections satisfies $R''>0$ at early times, and $R''\le0$ 
as $\tau\rightarrow\infty$, where the primes denote derivatives with respect
to the comoving time.  On the other hand, in the $k=1$ models the comoving
time runs within a finite range, and the scale factor $R$ diverges at both
ends of the interval.  We also obtained large classes of more complicated 
solutions whose cosmological properties we did not examine in detail, 
including those corresponding to the $SL(N+1,R)$ Toda equations.  It would 
be interesting to investigate the cosmology of these models further.

\section*{Acknowledgement}

K.-W.X.\ is grateful to TAMU for hospitality in the early stages of
this work.

\end{document}